\documentclass[final,1p,times]{elsarticle} % do not change
\usepackage{graphicx} % do not change
\usepackage{amssymb} % do not change
\usepackage{amsthm} % do not change
\usepackage{lineno} % do not change

\journal{Nuclear Physics A} % do not change
\begin{document} % do not change

\begin{frontmatter} % do not change

%% QM09Author: please enter your
%% Title, author and address info here; please do not use footnotes

% Your Title - please insert
\title{PHENIX Measurements of Azimuthal Anisotropy for $\pi^0$\\
Production at High $p_T$ in Au+Au Collisions at
$\sqrt{s_{NN}}=200$GeV}

% Principle author, and co-authors - please insert
\author{Rui Wei$^{a}$ for the PHENIX collaboration}

% Address - please insert
\address[a]{Chemistry Department, Stony Brook University, Stony Brook, NY 11794, USA}
\begin{abstract} % do not change
%% Text of abstract goes here - please insert
%
An improved measurement of $v_2$ for $\pi^0$ in a broad range of
$p_T$ and centrality is presented. By combining $v_2$ with the
$R_{AA}$, we provide new insights on jet-medium interactions.
We show that current pQCD energy loss models cannot describe the
suppression of the $\pi^0$ as a function of the angle with respect to
the reaction plane. Our result could help to resolve the factor of
4 differences in the predicted transport coefficients among these
models. Alternatively, it may suggest that non-perturbative effects
associated with the strongly coupled QGP are important, and new
theoretical developments are needed to fully understand the jet
medium interactions.

\end{abstract} % do not change

\end{frontmatter} % do not change

%% QM09: we keep linenumbers at least for initial version
%\linenumbers % do not change

%% start of main text - please insert.

%%\section{}\label{}

%\section{Introduction}
The measurement of $\pi^0$ anisotropy at PHENIX
potentially serves three important goals for heavy-ion program at
RHIC: a) to better constrain the energy loss processes by studying
the suppression of $\pi^0$ as function of angle to the reaction
plane (RP); b) to investigate the mechanisms for anisotropy at
intermediate $p_T$, where the transition from a pressure gradient
dependent outward-push driven by collective flow to a path length
dependent attenuation driven by jet quenching takes place; c) to
understand the non-flow contribution, especially those from jets,
in the high $p_T$ $v_2$ measurements. These goals are motivated by
the current limitation in our understanding of the interactions
between jets and flowing medium. For example, the extraction of
transport properties suffer from large theoretical uncertainties.
Predictions by various jet quenching frameworks differ by a factor of 4,
even though they all describe the $R_{AA}$ measurements with
identical 3-D hydro evolution of the underlying
medium~\cite{Bass:2008rv}. Precision measurements of RP dependent
suppression can provide severe constraints on these models.

These goals are facilitated by the large increase in the effective
statistics of the 2007 Au+Au dataset from PHENIX, which is the basis of
current analysis. It has two important improvements over
the previous PHENIX
measurements~\cite{Adler:2006bw,Afanasiev:2009iv}: 1) various new
RP detectors were installed covering a broad $\eta$ range with
improved resolution (RxNPin, RxNPout, MPC at $1.0<\eta<1.5$,
$1.5<\eta<2.8$ and $3<\eta<4$, respectively), and 2) factor of 4
increase in collected statistics relative
to~\cite{Afanasiev:2009iv}. This is equivalent to an effective
factor of 14 and 7 increase for $\pi^0$ $v_2$ measurement for RxNP
and MPC, respectively.  Due to limited space, we shall focus our
discussion on goal a) and briefly mention goals b) and c) at
relevant places.

%\section{Analysis}
The analysis follows the same technique laid out
in~\cite{Afanasiev:2009iv}. The main idea is to extract $\pi^0$'s
from the invariant mass distribution constructed separately in six
bins of $\pi^0$ angle relative to the event plane (EP),
$\Delta\phi=\phi_{\pi^0}-\Psi_{EP}$, in the interval
$\Delta\phi\in[0-\pi/2]$. The raw $v_2$ is obtained by fitting the
raw yield with $A(1+2v_2^{raw}\cos2\Delta\phi)$, which is then
corrected by the corresponding reaction plane resolution,
$\sigma_{RP}$, to give the signal $v_2 = v_2^{raw}/\sigma_{RP}$.
Note that the EP is obtained via standard event plane method and
$\sigma_{RP}$ is measured via the sub-event method by correlating
the north and south RP detectors (which are symmetric). The
differential nuclear modification factor $R_{AA}
(\Delta\phi_i,p_T)$ is calculated from the published angle-averaged
$R_{AA}(p_T)$~\cite{Adare:2008qa} as
\begin{equation}
\label{eq:1} R_{AA}(\Delta\phi_i,p_T) = R_{AA}(p_T)
\frac{N(\Delta\phi_i,p_T)}{<N(\Delta\phi,p_T)>}
\frac{1+2v_{2}\cos2\Delta\phi_i}{1+2v_{2}^{raw}\cos2\Delta\phi_i}.
\end{equation}

\begin{figure}[ht]
\centering
\begin{tabular}{lr}
\begin{minipage}{0.7\linewidth}
\begin{flushleft}
\includegraphics[width=0.96\linewidth]{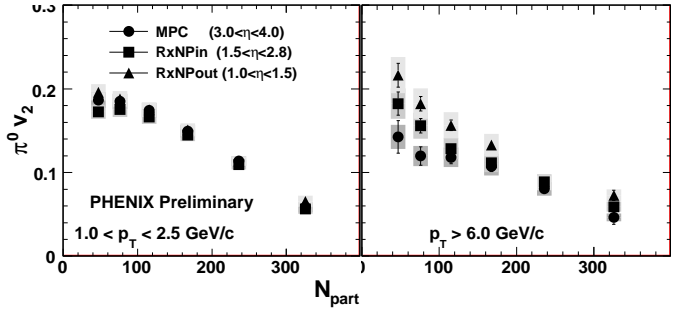}	
\end{flushleft}
\end{minipage}
&
\hspace*{0in}\begin{minipage}{0.20\linewidth}
\begin{flushright}
\caption{\label{fig:1} $v_2$ vs $N_{part}$ at a low $p_T$  and a
high $p_T$ regions using different reaction plane detectors.}
\end{flushright}
\end{minipage}
\end{tabular}
\end{figure}

%\section{Results}
One advantage of the broad $\eta$ coverage for the RP detectors is
that one can evaluate the potential non-flow effects.
Figure~\ref{fig:1} shows the $v_2$ vs $N_{part}$ in a low $p_T$
region and a high $p_T$ region, measured by different RP detectors.
There is a clear $\eta$ dependence of $v_2$ values at high $p_T$,
especially in the peripheral collisions. This is not the case at
low $p_T$. This suggests that most of the non-flow effects are
induced by jets, and they only become important at high $p_T$ and
for RP detectors reside closer to central arm where the $\pi^0$s
are detected. Even though RxNP detectors have the best resolution,
they may suffer jet bias. Instead, we use MPC for EP determination.
The MPC sits at the same $\eta$ range as BBC detectors (used in
previous analysis); however it has 40\% better resolution compare
to the BBC since it measures both neutral and charged particles,
and has higher granularity.

\begin{figure}[h]
\centering
\includegraphics[scale=0.6]{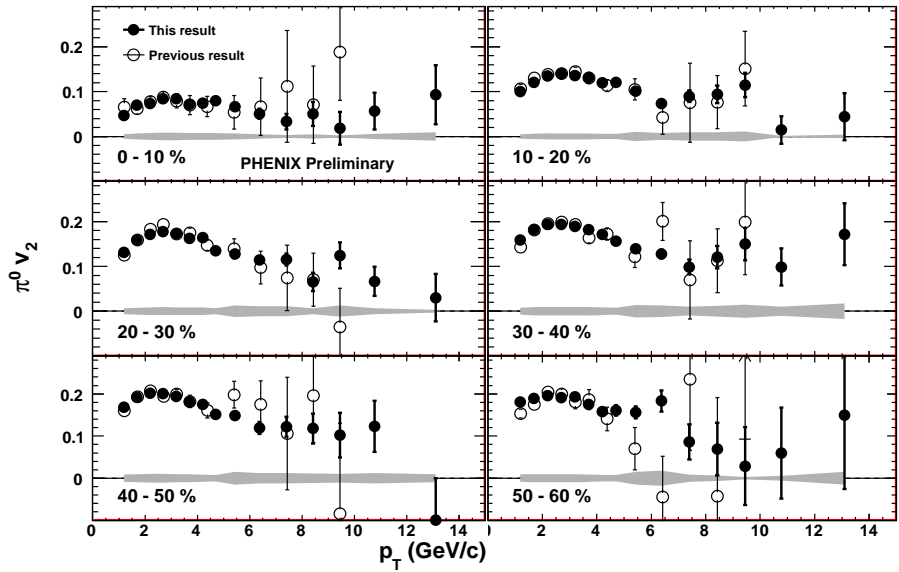}	
\caption{ \label{fig:2}
$\pi^0$ $v_2$ vs $p_T$ in six centrality bins. The solid points are the
PHENIX preliminary results in Run 7 measured with the MPC detector. The open points are the
published results from ~\cite{Afanasiev:2009iv}.}
\end{figure}

Figure~\ref{fig:2} shows $v_2$ values measured in six centrality
classes, overlaid with the published results from 2004 Au+Au
run~\cite{Afanasiev:2009iv}. The two measurements are consistent,
but new data improve significantly on both statistical errors and
$p_T$ reach. The $v_2$ at $p_T>6$ GeV/c, for all centralities,
remains significantly above zero and is constant with $p_T$.

\begin{figure}[th]
\includegraphics[scale=0.65]{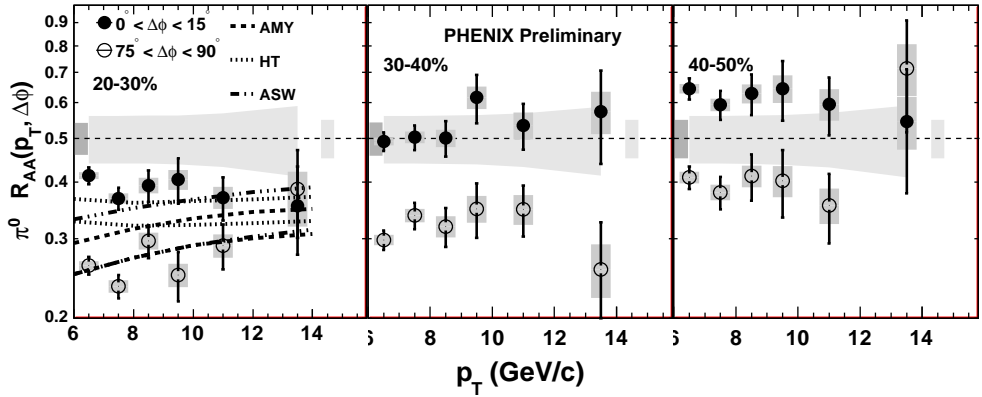}	
\caption{\label{fig:3} Data points: solid(open) is the
in-plane(out-of-plane) $\pi^0$ $R_{AA}(p_T)$ for three centrality bins measured with the MPC detector. Lines represent calculations of AMY, HT and ASW models from~\cite{Bass:2008rv}, respectively.}
\end{figure}

%, which can be used to constrain our understanding on the path length dependence of jet quenching
If the particle production at high $p_T$ ($>6$ GeV/c) is dominated
by fragmentation of the jets that survive the medium, as most
energy loss models predict, then it would lead to an anisotropy of
the suppression which can be compared with our data.
Figure~\ref{fig:3} shows the high $p_T$ $R_{AA}$ for $\pi^0$
measured in-plane ($0<\Delta\phi<\pi/12$) and out-of-plane
($5\pi/12<\Delta\phi<6\pi/12$) directions, derived according to
Eq.~\ref{eq:1}. A clear split of the suppression levels can be seen
between the two for three centrality bins shown (20-30\%, 30-40\%
and 40-50\%). Calculations from three pQCD jet quenching models
(abbreviated as AMY, HT, ASW) from~\cite{Bass:2008rv} are compared
with our data. These calculations are carried out with identical
initial conditions, medium evolution and fragmentation functions,
and their suppression levels have been tuned to reproduce the
inclusive $R_{AA}$ data in central Au+Au collisions as well as the
20-30\% bin. However, as Figure~\ref{fig:3}a shows, all three
models under-predict the difference between in-plane and
out-of-plane $R_{AA}$: The AMY model describes the out-of-plane
$R_{AA}$, but not the in-plane; HT model is the opposite; ASW model
does a reasonable job for both in- and out-of plane directions at
high $p_T$, but misses the low $p_T$ region. However, we are not
advocating that one model is better than the other, because
changing other control parameters, like initial conditions, medium
evolution or fragmentation functions, may change this comparison.
Our principal conclusion is that by utilizing both observables, one
should be able to narrow down the uncertainties on initial
condition and medium evolution and other effects that were ignored
so far in the calculation, e.g. virtuality difference between the
in- and out-of plane. Our data may help to resolve the
discrepancies, e.g. $\hat{q}$,  between the model predictions.

\begin{figure}[th]
\begin{tabular}{lr}
\begin{minipage}{0.6\linewidth}
\begin{flushleft}
\includegraphics[width=0.85\linewidth]{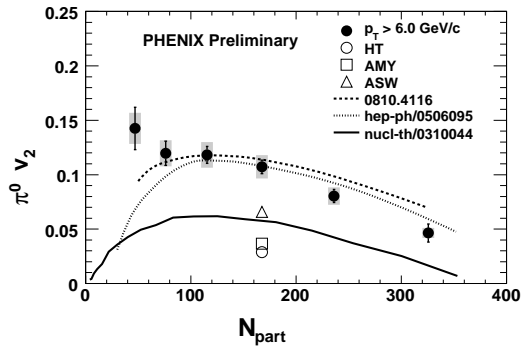}	
\end{flushleft}
\end{minipage}
&
\hspace*{0in}\begin{minipage}{0.30\linewidth}
\begin{flushright}
\caption{\label{fig:4} Data points: $v_2$ values at $p_T>$6GeV/$c$ measured with the MPC detector. Open symbols: pQCD model calculations taken
from~\cite{Bass:2008rv}. Lines: geometric model calculationss
with different assumptions on path length dependence
~\cite{Drees:2003zh,Pantuev:2005jt,Liao:2008dk}. }
\end{flushright}
\end{minipage}
\end{tabular}
\end{figure}

On the other hand, the failure of these models to reproduce the
large anisotropy at high $p_T$ may imply that the pQCD treatment of
the energy loss process as sequential radiation, being proportional
to the local color change density is not sufficient. In the
presence of the strongly coupled medium (sQGP), both the path
length dependence and the color charge dependence could be
modified. For example, calculation based on ADS/CFT technique
suggests that $\Delta E \propto L^3$~\cite{Gubser:2008as} and
$\hat{q} \propto \sqrt{\alpha_{SYM}N_c}$~\cite{Liu:2006ug} instead
of $\Delta E \propto L^2$ and $\hat{q} \propto \alpha_{s}N_c^2$ for
pQCD. In a separate calculation, Kharzeev~\cite{Kharzeev:2008qr}
estimates $dE/dx \propto E^2$ in strong coupling limit, much
stronger than the logarithmic dependence expected from pQCD. These
non-linear dependences could be the reason for the large
anisotropy. Figure~\ref{fig:4} shows the comparison of $v_2$ with
several toy model predictions with different medium density
dependence. The model from~\cite{Liao:2008dk} introduces strong
non-linear dependence by assuming that jet quenching happens mostly
close to phase transition boundary. The model
from~\cite{Pantuev:2005jt} does so by suppressing the energy loss
at high energy density region by using a plasma formation time
argument. Both are able to qualitatively describe centrality
dependence. However, much need to be done to generalize these toy
models into rigorous theoretical calculations.

In summary, we presented high statistics measurement of $\pi^0$
$v_2$ and $R_{AA} (\Delta\phi, p_T)$ up to 13 GeV/c in broad
centrality ranges of Au+Au collisions at $\sqrt{s_{NN}}$=200 GeV.
Current energy loss models were not able to describe
$R_{AA}(\Delta\phi, p_T)$ at $p_T>6$ GeV/c. This discrepancy
implies that the current pQCD models need further tuning, in which
case our measurement can help to resolve the differences in these
models. It may also suggest that the non-perturbative effects
associated with the strongly coupled QGP are important, and new
tools are required to understand the energy loss processes. Further
detailed study of the $v_2$ and $R_{AA}$ in the intermediate $p_T$ range
of 2-6 GeV/c may shed light on the role of perturbative and
non-perturbative effects.

%% end of main text

\section*{Acknowledgments} % please insert, comment out or delete if not needed
This work is supported by the NSF under award number PHY-0701487.

 % do not change
\end{document}